\documentstyle[aps,preprint,prc,12pt]{revtex}

\begin{document}

\title{
Circumstantial Evidence for a Critical Behavior in
Peripheral Au + Au
Collisions at 35 MeV/nucleon
}

\author{
P. F. Mastinu$^{1}$, M. Belkacem$^{1,9}$, M. D'Agostino$^1$,
M. Bruno$^1$, P. M. Milazzo$^{1,2}$, G. Vannini$^2$,
D. R. Bowman$^7$, N. Colonna$^3$,
J. D. Dinius$^6$,
A. Ferrero$^{4,8}$, M. L. Fiandri$^1$, C. K. Gelbke$^6$, T. Glasmacher$^6$,
F. Gramegna$^5$, D. O. Handzy$^6$, D. Horn$^7$, W. C. Hsi$^6$, M. Huang$^6$,
I. Iori$^4$, G. J. Kunde$^6$,
M. A. Lisa$^6$, W. G. Lynch$^6$,
G. V. Margagliotti$^2$, C. P. Montoya$^6$, A. Moroni$^4$, G. F. Peaslee$^6$,
F. Petruzzelli$^4$, L. Phair$^6$, R. Rui$^2$,
C. Schwarz$^6$, M. B. Tsang$^6$, C. Williams$^6$,
V. Latora$^9$ and A. Bonasera$^9$
}

\address{
$^{1}$ Dipartimento di Fisica and INFN, Bologna, Italy \\
$^{2}$ Dipartimento di Fisica and INFN, Trieste, Italy \\
$^{3}$ INFN, Bari, Italy \\
$^{4}$ Dipartimento di Fisica and INFN, Milano, Italy \\
$^{5}$ INFN, Laboratori Nazionali di Legnaro, Italy \\
$^{6}$ NSCL, Michigan State University, USA \\
$^{7}$ Chalk River Laboratories, Chalk River, Canada \\
$^{8}$ CNEA, Buenos Aires, Argentina \\
$^{9}$ INFN, laboratorio Nazionale del Sud, Catania, Italy
}

\date{ \today }
\maketitle


\vspace{0.2 cm}
\centerline{ (MULTICS - MINIBALL Collaboration) }
\vspace{0.2 cm}

\newpage

\begin{abstract}

The fragmentation resulting from peripheral Au + Au
collisions at an incident energy of E = 35 MeV/nucleon is
investigated. A power-law charge distribution, $A^{-\tau}$ with
$\tau \approx 2.2$, and an intermittency signal are observed for
events selected in the region of the Campi scatter plot
where "critical" behavior is expected.

\end{abstract}

{
\vskip 2\baselineskip
}

\newpage


Initiated by the observation of fragments in the final
stages of the reaction exhibiting a power law in fragment
charge distributions \cite{eos1}, and stimulated by the similarity
of the nuclear matter equation of state with that of a van
der Waals gas \cite{nm_eos}, the possibility of observing a liquid-gas
phase transition in nuclear systems has been the subject
of intensive investigations \cite{lg1,lg2,eos2,belkacem,aladin}.
This interest increased
recently with attempts of extracting critical exponents of
fragmenting nuclear systems produced in Au + C collisions
at E = 1 GeV/nucleon \cite{eos2} and a "caloric curve" for projectile
fragmentation reactions in Au + Au collisions at 600 MeV/nucleon \cite{aladin}.
In this letter, we report results obtained for
peripheral Au + Au collisions at E = 35 MeV/nucleon which
display some characteristics similar to those predicted for
near-critical systems.

The experiment was performed at the National
Superconducting Cyclotron Laboratory of the Michigan
State University. Fragments with charge up to $Z = 83$ were
detected at $3^o \le \theta_{lab} < 23^o$ by the Multics array \cite{strum},
and
charged particles with charge up to $Z = 20$ were detected
at $23^o \le \theta_{lab} \le 160^o$ by 159 phoswich detector elements
of the MSU Miniball \cite{mini}. The charge identification
thresholds were about 1.5 MeV/nucleon in the Multics
array, independently of the fragment charge, and about 2,
3, 4 MeV/nucleon in the Miniball for Z = 3, 10, 18,
respectively. The geometric acceptance of the combined
apparata was greater than 87\% of $4\pi$.

Our analysis is guided by calculations \cite{belkacem,gross,bondorf} which
predict that the "critical" excitation energy may decrease
when the system is either charged and/or rotating. Short-
lived systems formed in central Au + Au collisions are
predicted to expand and undergo a multi-fragment
breakup due to the high combined charge of projectile and
target nuclei. Evidence for such Coulomb driven breakup
of a single source has, indeed, been observed \cite{central} in
central collisions for this reaction. For larger impact
parameters, however, several smaller sources can emerge
corresponding to the decay of projectile and target-like
residues and a "neck" \cite{montoya} that momentarily joins them.
With an appropriate reaction filter, one might then hope
to select primary fragments with excitation energies,
Coulomb charges, and angular momenta appropriate to
bring the system into different portions of the instability
region \cite{gross,bonasera}.

Studies with the Classical Molecular Dynamics (CMD)
model indicated that critical behavior may be achieved for
peripheral Au + Au collisions at E = 35 MeV/nucleon \cite{theo}. When
these CMD results are filtered by the acceptance of the
Multics-Miniball arrays, the signals which may be
indicative of criticality become washed out due to poorly
detected events, but they can be recovered by restricting
the analysis to more completely detected peripheral
events for which the largest projectile-like fragment (PLF)
is detected. Guided by these calculations, we select events
for which the largest fragment has a velocity along the
beam axis larger than 75\% of the beam velocity and the
total detected charge is between 70 and 90. For these
events, the total detected linear momentum is larger than
50\% of the beam momentum. Their distribution of charged
particle multiplicities, $N_{c}$, is shown by the solid histogram
in Fig. 1.

Figure 2 shows a scatter plot of $ln(Z^{j}_{max})$ versus $ln(M^{j}_{2})$
("Campi scatter plot" \cite{campi1}) where $Z^{j}_{max}$
is the charge of the heaviest fragment and $M^{j}_{2}$
is the second conditional moment of the charge distribution
detected in the $j$-th event,
\begin{equation}
M^{(j)}_{2} = \sum_{Z} Z^{2} n_{j}(Z)
\end{equation}
Here, $n_{j}(Z)$ denotes the number of fragments of charge $Z$
detected in the $j$-th event, and the summation is over all
fragments but the heaviest detected one. Theoretical
investigations suggest that such plots may be useful in
characterizing near-critical behavior of finite systems \cite{campi1}.
The calculated Campi scatter plots typically exhibit two
branches: an upper branch with a negative slope
containing largely undercritical events (e.g. $T < T_{crit}$ in a
liquid-gas phase transition or $p > p_{crit}$ in a percolation
phase transition) and a lower branch with a positive slope
containing largely overcritical events ($T > T_{crit}$ or $p <
p_{crit}$). The two branches were shown to meet close to the
critical point of the phase transition \cite{belkacem,campi1,gross1}.

The data shown in Fig. 2, display two branches similar to
the ones predicted for undercritical and overcritical
events. In the top-right part, close to the intersection of
these two branches (potentially containing near-critical
events), a separate island is observed which is due to
fission events, as first noted by Ref. \cite{gross1}. By appropriate
gates in the Campi plot, these fission events are removed
from the following analysis.

To further investigate the two branches observed in Fig. 2
and the region where they intersect, we employ three cuts
selecting the upper branch (cut 1), the lower branch (cut
3) and the intersection region (cut 2); these cuts are
indicated in Fig. 2. The charged particle multiplicity
distributions observed for these three cuts are shown as
dashed histograms in Fig. 1. Cuts 1 and 3 largely select
low and high multiplicity events corresponding to very
peripheral and central collisions (assuming on the average
a monotonic
relation between $N_{c}$ and impact parameter); cut 2
represents a wide range of charged particle multiplicities
and thus may involve a wide range of intermediate impact
parameters. Thus emission from a unique source cannot
be ascertained for cut 2, and it is likely that this cut
contains contributions from projectile and target-like
sources and from the neck \cite{montoya} which emits lighter
fragments ($Z \approx 6 - 9$) with enhanced probability as
compared to the projectile residue \cite{montoya}.
However, one does not exclude that this large multiplicity distribution is
related to the occurrence of large fluctuations as expected at the critical
point.

Fragment charge distributions, not corrected for detection
efficiency, are presented \cite{note} for the three cuts in Fig. 3.
Cut 1 (crosses) contains both light fragments and heavy
residues and thus resembles the "U"-shaped distributions
predicted by percolation calculations in the sub-critical
region \cite{bauer}. For cut 3 (open circles), one observes an
unusually flat charge distribution similar to the one
previously reported \cite{central} for central collisions which were
selected without the specific constraints employed in this
paper and attributed to a Coulomb-driven breakup of a
very heavy composite system \cite{central} (The steep fall-off at
large $Z$ is an artifact of the selection of events with $Z = 70 - 90$
used in this paper). For cut 2 (solid points), a
fragment charge distribution is observed which resembles
a power-law distribution, $P(Z) \propto Z^{-\tau}$, with $\tau \approx 2.2$.
For macroscopic systems exhibiting a liquid-gas phase
transition, such a power-law distribution is predicted to
occur near the critical point \cite{fisher}. However, it is not yet
known by how much the final fragment distributions
differ from the primary ones after the sequential decays
of particle unstable primary fragments.

Figure 4 shows \cite{note} the logarithm of the scaled factorial
moments (SFM), defined as
\begin{equation}
F_i(\delta s)={{\sum _{k=1}^{Z_{tot}/ \delta s}<{n_k}\cdot ({n_k}-1)\cdot ...
\cdot({n_k}-i+1)>}
\over {\sum _{k=1}^{Z_{tot}/ \delta s}<n_k>^i}}
\label{SFM}
\end{equation}
($i$ = 2, ..., 5), as a function of the logarithm of the bin size
$\delta s$. In the above definition of the SFM, $Z_{tot} = 158$ and $i$ is
the order of the moment. The total interval $[1, Z_{tot}]$ is
divided into $M = Z_{tot}/\delta s$ bins of size $\delta s$, $n_k$ is the
number
of particles in the $k$-th bin for an event, and the brackets $< >$
denote the average over many events. A linear rise of
the logarithm of the SFM versus $\delta s$ (i.e.
$F_{i} \propto \delta s^{-\lambda_{i}}$) indicates an
intermittent pattern of fluctuations \cite{gross1,bialas,plocia}. Even
though this quantity is ill defined for fragment
distributions \cite{campi2,phair,delzoppo}, several theoretical studies have
indicated that critical events give a power law for the SFM
versus the bin size \cite{belkacem,plocia}. For cut 3 (right part of the
figure), the logarithm of the scaled factorial moments
$ln(F_{i})$ is always negative (i.e. the variances are smaller
than Poissonian \cite{plocia}) and almost independent on $\delta s$; there
is no intermittency signal. The situation is different for cut
2 (central part). The logarithm of the scaled factorial
moments are positive and almost linearly increasing as a
function of $-ln(\delta s)$, and an intermittency signal is observed.
Region 1, corresponding to evaporation, gives zero slope.
Increasing or reducing the size of the three cuts in the
respective regions does not change significantly these
results. The interpretation of experimentally observed
intermittency signals may, however, be problematic due
to ensemble averaging effects \cite{phair}, even though calculations show that
impact parameter averaging only increases the absolute value of the SFM
\cite{delzoppo}.
Since cut 2 may involve a
large range of impact parameters, the observed
intermittency signal could be an artifact of ensemble
averaging and can, therefore, not be taken as a definitive
proof of unusually large fluctuations in a sharply defined
class of events.

In conclusion, we have analyzed fragment production in
Au + Au collisions at E = 35 MeV/nucleon. Events were selected
by requiring a total detected charge between 70 and 90
and the velocity of the largest detected fragment larger
than 75\% of the projectile velocity. A Campi scatter plot of
these events displays two branches similar to the sub-
and overcritical branches observed in theoretical studies.
The selection of events from the intersection of these two
branches (which has been associated with critical events
in theoretical studies) shows a power law charge
distributions with an exponent of $\tau\approx 2.2$ similar to that
characterizing the mass distribution near the critical point
of a liquid-gas transition. These events, further, display
an intermittent behavior similar to that expected for
near-critical events. While these signatures have been
associated with near-critical events, we must caution that
the effects of finite experimental acceptance and event
mixing with possible contributions from the decay of
projectile-like fragments and the neck-region are not yet
sufficiently well understood to allow an unambiguous
conclusion of critical behavior in the present reaction. Our
work does, however, show that different regions of the
nuclear phase diagram can be probed at one incident
beam energy by selecting events according to different
impact parameters and/or energy depositions.

One of the authors (M. Belkacem) thanks the Physics
Department of the University of Trieste for financial
support and the Physics Department of the University of
Bologna, where the major part of this work has been done,
for warm hospitality and financial support. This work has
been supported in part by funds of the Italian Ministry of
University and Scientific Research. The authors would like
to thank R. Bassini, C. Boiano, S. Brambilla, G. Busacchi, A.
Cortesi, M. Malatesta and R. Scardaoni for their technical
assistance.


\newpage

\begin{figure}
\label{f1}
\noindent
\caption{Multiplicity distribution of the
events selected for the analysis (solid histogram). The three dashed
histograms (1, 2 and 3) represent the multiplicity
distributions of the events falling in the three cuts drawn in Fig. 2.}
\end{figure}

\begin{figure}
\label{f2}
\noindent
\caption{Campi scatter plot, $ln(Z_{max})$ versus $ln(M_{2})$.
The three different regions are discussed in the text. Fission events are
to the right of region 2.}
\end{figure}

\begin{figure}
\label{f3}
\noindent
\caption{Charge distributions for the three cuts indicated in Fig. 2. The
curve represents a power law distribution $Z^{-\tau}$ with $\tau = 2.2$.}
\end{figure}

\begin{figure}
\label{f4}
\noindent
\caption{Scaled factorial moments
$ln(F_{i})$ versus $- ln(\delta s)$ for the three cuts indicated in Fig. 2;
the left part of the figure corresponds to cut 1, the
central part to cut 2 and the right part to cut 3. Solid points represent the
SFM of order
$i=2$, open circles
$i=3$, open squares $i=4$, and open triangles $i=5$.}
\end{figure}


\begin{thebibliography}{99}
\bibitem{note} The charge distributions shown in Fig. 3 and
the SFM shown in Fig. 4 have been calculated without the heaviest fragment
in each event, apart the charge distribution of cut 1.
\end{thebibliography}
\end{document}